# Teaching special relativity at the hundredth anniversary of Einstein's "anno mirabilis"

Bernhard Rothenstein
"Politehnica" University of Timisoara, Physics Department
Timisoara, Romania

## 1. The teaching strategy

Consider that at a given day of the year 2005 an instructor starts, at the request of his students, to introduce them in the fundamental problems of special relativity. He knows that since 1905, when Einstein submitted his first paper on special relativity "On the electrodynamics of moving bodies"[1] university-level introductory physics and modern physics textbooks and some higher level books present the subject.[2,3,4] They present the problems are in a similar way becoming a tradition in teaching strategy. Journals devoted to physics teaching publish an increasing number of papers that propose transparent approaches in order to ease the implementation of anti-common sense effects in the mind of the learners. The instructor will consider them as a tribute for all that they have done.

In order to be fair with his audience the instructor states that his students should be able to define the classical meaning of the physical quantities introduced in the different chapters of physics (kinematics, dynamics and electricity) and to know how we measure them. The knowledge of classical (Galilean) relativity is helpful. The mathematics he uses is simple.

Mathematics accustoms the students with postulates and so the instructor will not surprise them telling that all the experiments physicists perform should lead to results in accordance with the mighty principle of relativity that can be expressed in the form of two postulates: (1) The laws of physics are the same in all inertial reference frames. (2) The speed of light in vacuum has the same numerical value $c$ when measured in any inertial reference frame, independent of the motion of the source of light and/or observer.

Special relativity becomes involved when observers from two inertial reference frames in relative motion measure one and the same physical property of an object. The object can be at rest in one of the reference frames but moving relative to the other or can move relative to both of them. The fundamental problem is to find out a relationship between the results they obtain.

The involved reference frames are K(XOY) and K'(X'O'Y'). The corresponding axes of the two frames are parallel to each other. The OX(O'X') axes are overlapped and K' moves relative to K in the positive direction of the overlapped axes with constant velocity $V$. At the common origin of time the origins O and O' of the two frames are shortly located at the same point in space. We present a physical quantity measured by observers from K as unprimed whereas physical quantities measured by observers from K as primed.

The following assumptions have deep roots in the mind of the audience:
-lengths of objects are the same in all inertial reference frames,
-time passes at the same rate in different reference frames,
-measuring the inertial mass of a body (its resistance to acceleration) observers from K and K' obtain the same value.

The instructor mentions that the assumptions are not in accordance nor with experiment neither with the principles of relativity and invites the audience to delete them from memory. He is aware of the fact that the lecture should be time saving because many colleges and even universities avoid to include the teaching of special relativity invoking lack of time.

Physicists operate with basic physical quantities (length, time, mass and electric charge) and with combinations of them. For each of them they define the concept of proper value. The proper length of a rod is its length measured by an observer relative to whom it is in a state of rest. We measure the proper time interval between two events measured as a difference between the readings of the same clock in front of which the two events take place. The observer relative to whom a particle is in a state of rest measures its proper mass. The concept of electric charge has a special status, its magnitude being the same in all inertial reference frames.

An equation that relates a physical quantity measured in K to the same physical quantity measured in K' represents a transformation equation. The simplest transformation equation relates a proper physical quantity measured say in K' to the same physical quantity measured in K. In that case the physical object on which measurements are performed is at rest in K' but moves with velocity $V$ relative to K.

Thought experiments (and effectively performed ones) interpreted in terms of the principle of relativity lead to the following simple transformation equations

$$L = L_0 \sqrt{1 - \frac{V^2}{c^2}} = \gamma^{-1}(V) \tag{1}$$

$$t = \frac{t_0}{\sqrt{1 - \frac{V^2}{c^2}}} = \gamma(V) t_0 \tag{2}$$

$$m = \frac{m_0}{\sqrt{1 - \frac{V^2}{c^2}}} = \gamma(V) m_0 \tag{3}$$

It is important to know that:

-$L_0$ represents the proper length of a rod at rest in K' and located along the overlapped axes OX(O'X'), $L$ representing the length of the same rod measured by observers from K,

-$t_0$ represents the reading of a clock $C'_0(0,0)$ when it is shortly located in front of a clock $C(x,0)$ of the K frame that reads $t$ but reads a zero time when it is shortly located in front of a clock $C_0(0,0)$ of the K frame that reads a zero time as well.

-$m_0$ represents the rest mass of a particle measured by observers from K', $m$ representing its mass measured by observers from K.

Because $\gamma(V) < 1$ relativists say that length shrinks, time dilates and mass increases.

The principle of reciprocity tells us that
-if you move relative to me with velocity **u** I move relative to you with velocity –**u,**
-if you measure a proper physical quantity and you move relative to me, I measure its contracted or dilated value.

The scenario we follow, involves a particle that moves with velocity **u**$(u_x, u_y)$ relative to K and with velocity **u'** $(u'_x, u'_y)$ relative to K'. The particle goes at the origin of time in the two frames ($t=t'=0$), through the overlapped origins O and O' of the two frames, and moves along a direction that makes an angle $\theta$ when detected from K and an angle $\theta'$ when detected from K', both with the positive direction of the overlapped axes OX(O'X'). It is obvious that

$$u_x = u\cos\theta \tag{4}$$
$$u_y = u\sin\theta \tag{5}$$
$$u'_x = u'\cos\theta' \tag{6}$$
$$u'_y = u'\sin\theta'. \tag{7}$$

The velocities **u** and **u'** have a radial character in K and K' respectively. Thought experiments lead to the following relativistic velocity transformations[5] without sing the Lorentz-Einstein transformations of the space-time coordinates of the same event

$$u_x = \frac{u'_x + V}{1 + \frac{Vu'_x}{c^2}} = u' \frac{\cos\theta' + \frac{V}{u'}}{1 + \frac{Vu'}{c^2}\cos\theta'} \tag{8}$$

$$u_y = \frac{u'_y\sqrt{1 - \frac{V^2}{c^2}}}{1 + \frac{Vu'_x}{c^2}} = \frac{u'\sqrt{1 - \frac{V^2}{c^2}}\sin\theta'}{1 + \frac{Vu'}{c^2}\cos\theta'}. \tag{9}$$

The magnitude of the velocity transforms as

$$u = u' \frac{\sqrt{(\cos\theta' + \frac{V}{u'})^2 + \left(1 - \frac{V^2}{c^2}\right)\sin^2\theta'}}{1 + \frac{Vu'}{c^2}\cos\theta'}. \tag{10}$$

The angles defined above transform as

$$\sin\theta = \frac{\sqrt{1 - \frac{V^2}{c^2}}\sin\theta'}{\sqrt{(\cos\theta' + \frac{V}{u'})^2 + (1 - \frac{V^2}{c^2})\sin^2\theta'}} \tag{11}$$

$$\cos\theta = \frac{\cos\theta' + \frac{V}{u'}}{\sqrt{(\cos\theta' + \frac{V}{u'})^2 + (1 - \frac{V^2}{c^2})^2\sin^2\theta'}}. \tag{12}$$

## 2. The Lorentz-Einstein transformations for the space-time coordinate of the same event

The concept of event is fundamental in special relativity. An event is a physical occurrence that could take place at a given point in space at a given time. We characterize an event by the space coordinates of the point where it takes place using Cartesian *(x,y,x'y')* or polar *(r,θ)* coordinates and by its time coordinate that equates the reading of the clock located where the event takes place. The clocks located at the different points of a reference frame display the same running time because observers from the corresponding reference frames synchronized them following a procedure proposed by Einstein[6]. The notations $E(x = r\cos\theta, y = r\sin\theta, t)$ in K and $E'(x' = r'\sin\theta', y' = r'\sin\theta', t')$ define events that take place in K and in K' respectively. We say that they represent the same event if they take place at the same point in space when the clocks $C(x = r\cos\theta, y = r\sin\theta)$ and $C'(x' = r'\cos\theta', y' = r'\sin\theta')$ shortly located at that point read $t$ and $t'$ respectively.

The particle that plays an important part in our scenario generates after a time $t(t')$ of motion the events $E(x = r\cos\theta, y = r\sin\theta, t = \frac{r}{u})$ and $E'(x' = r'\sin\theta', y' = r'\sin\theta', t' = \frac{r'}{u'})$ in K and in K' respectively. It is obvious that

$$u = \frac{r}{t} \tag{13}$$

$$u' = \frac{r'}{t'}. \tag{14}$$

Combining (13) and (14) and taking into account (10) we obtain

$$\frac{r}{t} = \frac{r'}{t'}\frac{u}{u'} = \frac{r'}{t'} \cdot \frac{\sqrt{(\cos\theta' + \frac{V}{u'})^2 + (1 - \frac{V^2}{c^2})\sin^2\theta'}}{1 + \frac{Vu'}{c^2}\cos\theta'}. \tag{15}$$

Eq.(14) suggests to consider that

$$r = f(V)r'\sqrt{(\cos\theta' + \frac{V}{u'})^2 + (1 - \frac{V^2}{c^2})\sin^2\theta'} \tag{16}$$

and that

$$t = f(V)t'(1 + \frac{Vu'}{c^2}\cos\theta') \tag{17}$$

where *f(V)* represents an unknown function of the relative velocity but not of the physical quantities involved in the transformation process, because to one event in K corresponds a single event in K' and vice-versa. The simplest way to find out *f(V)* is to impose that (17) accounts for time dilation i.e. for *u'=0* (17) should lead to (2) and so

$$f(V) = \frac{1}{\sqrt{1 - \frac{V^2}{c^2}}} = \gamma(V). \tag{18}$$

The transformation equations we are looking for become

$$r = r'\gamma(V)\sqrt{(\cos\theta' + \frac{V}{u'})^2 + (1 - \frac{V^2}{c^2})\sin^2\theta'} \tag{19}$$

$$t = \gamma(V)t'(1 + \frac{Vu'}{c^2}\cos\theta') = \gamma(V)(t' + \frac{V}{c^2}x'). \tag{20}$$

The Cartesian coordinates transform as

$$x = r\cos\theta = \gamma(V)r'(\cos\theta' + \frac{V}{u'}) = \gamma(V)(x' + Vt') \tag{21}$$

$$y = r\sin\theta = r'\sin\theta'. \tag{22}$$

The important result is that distances measured perpendicular to the direction of relative motion are the same when measured from different inertial reference frames. Thought experiments that do not use light signals prove that this is a direct consequence of the first postulate.[7]

**3. Relativistic dynamics**

Classical mechanics operates with concepts like mass, momentum, kinetic energy and energy. The problem is to find out theirs relativistic counterparts and the relationship between them when they are measured by observers from K and K' respectively. We follow the same scenario considering that our particle caries mass and momentum. The observers from K will consider that the momentum of the particle is

$$\mathbf{p} = m\mathbf{u} \qquad (23)$$

whereas observers from K' will consider that it is

$$\mathbf{p}' = m'\mathbf{u}'. \qquad (24)$$

Both know that $u$ and $u'$ are now relativistic velocities and so $m$ and $m'$ are not classical masses. Combining (23) and (24) and taking into account t (10) we obtain

$$\frac{p}{m} = \frac{p'}{m'}\frac{u}{u'}\frac{\sqrt{(\cos\theta' + \frac{V}{u'})^2 + (1-\frac{V^2}{c^2})\sin^2\theta'}}{1+\frac{Vu'}{c^2}\cos\theta'}. \qquad (25)$$

Eq.(25) suggests to consider that

$$p = f(V)p'\sqrt{(\cos\theta' + \frac{V}{u'})^2 + (1-\frac{V^2}{c^2})\sin^2\theta'} \qquad (26)$$

$$m = f(V)m'(1+\frac{Vu'}{c^2}\cos\theta'). \qquad (27)$$

We obtain function $f(V)$ imposing the condition that for $u'=0$ observers from K' measure the rest mass of the particle and (27) should account for the relativistic effect described by (3). The result is

$$f(V) = \gamma(V) = \frac{1}{\sqrt{1-\frac{V^2}{c^2}}}. \qquad (28)$$

The velocities $u$ and $u'$ having a radial character and the mass having a scalar character, the momentum of the particle has a radial character in K and in K' as well and so its components transform as

$$p_x = p\cos\theta = \gamma(V)p'(\cos\theta' + \frac{V}{u'}) = \gamma(V)(p'_x + Vm') \qquad (29)$$

$$p_y = p\sin\theta = p'\sin\theta' = p'_y \qquad (30)$$

resulting that the OY(O'Y') component of the momentum is a relativistic invariant.

Textbooks derive the transformation equations for mass and momentum considering collisions between two particles from K and K' and imposing the condition that conservation of mass and momentum should work in both of them.[8]

Relativists call $m_0$ rest (Newtonian) mass and $m$ relativistic mass. We find that notations in many textbooks Some Authors ban the concept of relativistic mass whereas others defend it. We can avoid the concept of relativistic mass multiplying both sides of (27) by $mc^2$. Doing so it becomes

$$mc^2 = \gamma(V)(m'c^2 + \frac{V}{c^2}m'u'_x). \qquad (31)$$

Physicists as experienced godfathers find out a name for the physical quantity $mc^2$ ($m'c^2$). Having the physical dimensions of energy it receives the name of relativistic energy $E(E')$

$$E = mc^2 \tag{32}$$
$$E' = m'c^2. \tag{33}$$

With the new notations (31) becomes

$$E = \gamma(V)(E' + \frac{V}{c^2} p'_x) \tag{34}$$

whereas (29) becomes

$$p_x = \gamma(V)(p'_x + \frac{V}{c^2} E'). \tag{35}$$

If m$(m')$ is a measure of the inertia of a particle then $E(E')$ is a measure of it as well. If the particle is at rest in K' observers of that frame measure its rest mass $m_0$ and its rest energy $E_0 = m_o c^2$ a concept that has no equivalent in classical physics. Because the single supplementary energy a free particle can possess is its kinetic energy $E_k$ we obtain for it

$$E_k = E - E_0 = m_0 c^2 (\frac{1}{\sqrt{1 - \frac{V^2}{c^2}}} - 1) = E_0(\gamma(V) - 1). \tag{36}$$

We do not enter in the philosophical interpretations of the mass-energy equivalence (mass and energy account for the same physical property of particles: theirs inertia) but we mention instead the following important consequences:

-a change in the rest mass of a particle entails a change in its rest energy,

-characterizing the dynamic properties of a particle we will say that it has a given rest energy $E_0 = m_0 c^2$. When it moves with velocity $\mathbf{u}(u_x, u_y)$ relative to K its energy becomes

$$E = \frac{E_0}{\sqrt{1 - \frac{u^2}{c^2}}} \tag{37}$$

whereas if it moves with velocity $\mathbf{u}(u'_x, u'_y)$ relative to K' its energy measured by observers of that frame becomes

$$E' = \frac{E_0}{\sqrt{1 - \frac{u^2}{c^2}}}. \tag{38}$$

(34) relating the two energies. Considering a collision between two particles from K and K', we obtain that applying the transformation equations derived above we obtain results in accordance with the conservation laws of mass (energy) and momentum.

## 4. Electrodynamics

As we have mentioned so far the electric charge has the same magnitude for all inertial observers. The scenario w follow involves a very large conducting surface confined in the $(X^0Y^0Z^0)$ plane defined by the axes of the $K^0$ reference frame, that moves with velocity $u_x$ relative to K and with velocity $u_x^{'}$ relative to K'. It generates an uniform electric field that shows in the positive direction of the $O^0Y^0$ axis given by[9]

$$E_y^0 = \frac{Q}{2\varepsilon_0 A^0} = \frac{Q}{2\varepsilon_0 l_x^0 l_z^0} \tag{39}$$

where Q is the invariant positive charge uniformly distributed on the surface $A^0 = l_x^0 l_z^0$ and $\varepsilon_o$ represents the vacuum permittivity, a relativistic invariant. Detected from K the same electric field is

$$E_y = \frac{Q}{2\varepsilon_o l_x^o \sqrt{1 - \frac{u_x^2}{c^2}}} \tag{40}$$

whereas detected from K' it is

$$E_y^{'} = \frac{Q}{2\varepsilon_o l_x^o \sqrt{1 - \frac{u_x^{'}}{c^2}} l_z^o} \tag{41}$$

because of the invariance of distances measured perpendicular to the direction of relative motion. Combining (40) and (41) we obtain

$$E_y = E_y^{'} \frac{\sqrt{1 - \frac{u_x^{'2}}{c^2}}}{\sqrt{1 - \frac{u_x^2}{c^2}}} = \gamma(V) E_y^{'} (1 + \frac{V}{c^2} u_x^{'}) = \gamma(V)(E_y^{'} + \frac{V}{c^2} u_x^{'} E_y^{'}) \tag{42}$$

that is a genuine transformation equation presenting in its left hand side a physical quantity measured in K whereas in its right hand side only physical quantities measured in K. We detect in the right hand side of (42) the presence of the term $\frac{u_x^{'}}{c^2} E_y^{'}$. Again, the physicist as a well-trained godfather introduces the notation

$$B_z^{'} = \frac{u_x^{'}}{c^2} E_y^{'} \tag{43}$$

and call that combination of physical quantities "the O'Z' component of the magnetic induction **B**'in such a way that the vector product

$$\mathbf{u}_x^{'} \times \mathbf{E}_y^{'} \tag{44}$$

shows in the positive direction of the O'Z'axis. The OZ component of the magnetic induction transforms as

$$B_z = \frac{u_x}{c^2}E_y = \gamma(V)\frac{E_y'}{c^2}u_x'(1+\frac{V}{u_x'}) = \gamma(V)B_z'(1+\frac{V}{u_x'}) = \gamma(V)(B_z' + \frac{V}{c^2}E_y') \qquad (45)$$

Confining the uniformly charged plane we have considered so far in the $X^0O^0Y^0$ plane it generates an uniform electric field $E_z^0$ that shows in the positive direction of the $O^0Z^0$ axis and an uniform magnetic field showing in the negative direction of the OY(O'Y') axis. Following the same strategy as above, we obtain the following transformation equations

$$E_z = \gamma(V)(E_z' - VB_y') \qquad (46)$$

$$B_y = \gamma(V)(B_y' - \frac{V}{c^2}E_z'). \qquad (47)$$

The traditional way to derive the transformation equations for the components of **E** and **B** uses the transformation of relativistic force[10]. Our derivation avoids its use takes into account the invariance of the electric charge and shows that a magnetic field is no more then a moving electric field. No reference is made to Maxwell's equations.

## 5. Generalizing the approach

The derivations in the previous sections generated the following transformation factors

-For transforming $x \to x'$ and $p_x \to p_x'$

$$D_{x \to x', p_x \to p_x'} = \gamma(V)(\cos\theta' + \frac{V}{u'}). \qquad (48)$$

-For transforming $y \to y'$ and $p_y \to p_y'$

$$D_{y \to y', p_y \to p_y'} = 1 \qquad (49)$$

-For transforming $r \to r', p \to p'$

$$D_{r \to r', p \to p'} = \gamma(V)\sqrt{(\cos\theta' + \frac{V}{u'}) + (1 - \frac{V^2}{c^2})\sin^2\theta'} \qquad (50)$$

-For transforming $t \to t', m \to m', E \to E'$

$$D_{t \to t', m \to m', E \to E'} = \gamma(V)(1 + \frac{Vu'}{c^2}\cos\theta'). \quad (51)$$

The transformation factors (50) and (51) have in common the fact that when the considered events are generated by the front of a plane electromagnetic wave ($u = u' = c$) they become equal to each other i.e.

$$D^c_{r \to r', p \to p'} = D^c_{t \to t', m \to m', E \to E'} = \gamma(V)(1 + \frac{V}{c}\cos\theta'_c) \quad (52)$$

resulting that the momentum and the energy carried by the electromagnetic wave transform as

$$p_c = \gamma(V)(1 + \frac{V}{c}\cos\theta') \quad (53)$$

$$E_c = \gamma(V)E'_c(1 + \frac{V}{c}\cos\theta'_c). \quad (54)$$

The directions along which the same light (photon) propagates when detected from K and K' respectively transform as

$$\cos\theta_c = \frac{\cos\theta'_c + \frac{V}{c}}{1 + \frac{V}{c}\cos\theta'_c}. \quad (55)$$

It is Einstein's merit to recognize that light behaves as discontinuous individual particles-photons. The energy $\varepsilon$ of such a light particle is proportional to the frequency $v$ of the electromagnetic oscillations taking place in the light wave

$$\varepsilon = hv \quad (56)$$

where $h$ represents Planck's constant. If the invariant number N of photons carries the energy Ec (54) then the energy carried by a single photon is

$$\varepsilon = \frac{E_c}{N} = hv \quad (57)$$

in K and

$$\varepsilon' = \frac{E'_c}{N} = hv' \quad (58)$$

in K' resulting that the frequency of the electromagnetic oscillations taking place in the wave transforms as

$$v = v'\frac{E_c}{E'_c} = \gamma(V)v'(1 + \frac{V}{c}\cos\theta'_c) = \gamma(V)(v' + V\frac{v'}{c}\cos\theta'_c). \quad (59)$$

We detect in the right hand side of (59) the presence of the term

$$\frac{v'}{c} = \frac{1}{\lambda'} = k'. \tag{60}$$

Relativists call it wave vector

$$\mathbf{k}' = \frac{v'}{c}\mathbf{n}' = \frac{1}{\lambda'}\mathbf{n}' \tag{61}$$

*n'* representing the unit vector that shows in the direction in which the light (electromagnetic) wave propagates, $\lambda'$ representing the wave length. The OX(O'X') component of the wave vector is

$$k'_x = k' \cos\theta'_c \tag{62}$$

in K' and

$$k_x = k \cos\theta_c \tag{63}$$

in K transforming as

$$k_x = \frac{v}{c}\cos\theta_c = \gamma(V)k'(\cos\theta'_c + \frac{V}{c}) = \gamma(V)(k'_x + \frac{V}{c^2}v'). \tag{64}$$

It is a simple exercise to show that the OY component of the wave vector is a relativistic invariant. Eq.(61) can be now presented as

$$v = \gamma(V)(v' + Vk_x). \tag{65}$$

Because deriving the transformation equations for **E** and **B** only velocities that show in the positive direction of the overlapped axes were considered, the corresponding transformation factors are (42),(45)

$$D_{E_y \to E'_y} = \gamma(V)(1 + \frac{V}{c^2}u'_x) \tag{66}$$

$$D_{B_z \to B'_z} = \gamma(V)(1 + \frac{V}{u'_x}). \tag{67}$$

Again for $u = u' = c$ the two transformation factors become equal to each other and so the in an electromagnetic wave the electric and the magnetic components transform as

$$E^c_y = E'^c_y \sqrt{\frac{1 + \frac{V}{c}}{1 - \frac{V}{c}}}$$

$$B^c_z = B'^c_z \sqrt{\frac{1 + \frac{V}{c}}{1 - \frac{V}{c}}}. \tag{68}$$

Traditionally, Authors derive (65) and (67) imposing the invariance of the phase of a plane electromagnetic wave a laborious, time consuming and not very transparent approach.[11] Our approach involves only the transformation of relativistic velocities and the fact that there is a reference frame in which we can define a proper value of the electric field intensity and does not make reference to Maxwell's equations.

## 6. Generating four vectors

Consider the obvious identity

$$\frac{1-\frac{u^2}{c^2}}{1-\frac{u^2}{c^2}} = 1 \tag{69}$$

in the K frame. Multiplying both its sides with $c^2\tau^2$ where $\tau$ represents a proper time interval (69) leads to

$$\frac{c^2\tau^2}{1-\frac{u^2}{c^2}} - \frac{u^2\tau^2}{1-\frac{u^2}{c^2}} = c^2\tau^2. \tag{70}$$

With the notations introduced above (79) becomes

$$c^2(\Delta t)^2 - r^2 = c^2\tau^2. \tag{71}$$

Because $c^2\tau^2$ is a relativistic invariant

$$c^2(\Delta t)^2 - r^2 = c^2(\Delta t)^2 - x^2 - y^2 - z^2 \tag{72}$$

is a relativistic invariant as well. We can prove that performing the transformation of (72) to K' and showing that it becomes

$$c^2(\Delta t')^2 - r'^2 = c^2(\Delta t')^2 - x'^2 - y'^2 - z'^2. \tag{73}$$

Relativists consider that $(c\Delta t), x, y, z$ are the components of a four vector, the components of which transform via the Lorentz-Einstein transformations for the space-time coordinates of the same event.

Multiplying both sides of (69) with the invariant expression $m_0^2 c^4$ it becomes

$$\frac{m_0^2 c^4}{1-\frac{u^2}{c^2}} - \frac{m_0^2 c^4 u^2}{1-\frac{u^2}{c^2}} = m_0^2 c^4. \tag{74}$$

With the notations introduced above (74) becomes

$$\frac{E^2}{c^2} - p^2 = \frac{E^2}{c^2} - p_x^2 - p_y^2 - p_z^2 = m_0^2 c^2. \tag{75}$$

Performing the transformation to K' frame via the corresponding transformation equations for energy and momentum it leads to

$$\frac{E'^2}{c^2} - p'^2_x - p'^2_y - p'^2_z = m_0^2 c^2 \tag{76}$$

proving the invariance of $\frac{E^2}{c^2} - p^2$. Relativists consider that $\frac{E}{c}, p_x, p_y, p_z$ are the components of a four vector that transform via the transformation equations for energy and momentum.

## 7. Conclusions

We have shown that the transformation of relativistic velocities that can be derived without using the Lorentz-Einstein transformations and the fact that in each chapter in which physics is divided we can define proper (invariant) physical quantities, lead to the transformation of
-space-time coordinates of the same event generated by a moving tardyon, light signal or photon,
-momentum-mass(energy) of the same tardyon(photon),
-electric field intensity-magnetic induction generated by the same moving charged particle or by an electromagnetic wave,
-frequency-wave number of the same electromagnetic wave.

      Most textbooks present transformation equations that contain in theirs left hand side a given physical quantity measured say in K whereas in theirs left hand side they contain the same physical quantity and another one both measured in K'. The transformation equations we have derived above present in theirs left hand side a given physical quantity measured in K whereas in theirs left hand side they contain the same physical quantity measured in K' and a transformation factor that depends on the relative velocity *V* and on the velocity of the considered particle relative to K'. The transformation factor that performs the transformation of the length of the position vector and of the magnitude of the momentum is equal to each other. The transformation factors that perform the transformation of time, mass (energy) are equal to each other as well. If the considered events are generated by a light signal or by a photon then the transformation factors that perform the transformation of the length of the position vector, of the time coordinate of the momentum and the energy carried by a photon are equal to each other.

      Involving a single scenario (a moving particle that generates events, carries mass (energy), electric charge or a light signal (photon), our approach is time saving, offering a correct grasp of all concepts relativists use.

      The section devoted to relativistic dynamics offers solutions for those who use the concept of relativistic mass but also for those who ban it. Each section is independent from the other ones.

      Physicists and engineers who use formulas in order to describe quantitatively relativistic effects should have derived them at least once. Our approach offers them that opportunity.